# Citation indices and dimensional homogeneity

**Gangan Prathap**

*A P J Abdul Kalam Technological University, Thiruvananthapuram, Kerala, India 695016. e-mail: gangan_prathap@hotmail.com*

**Abstract**

The importance of dimensional analysis and dimensional homogeneity in bibliometric studies is always overlooked. In this paper, we look at this issue systematically and show that most *h*-type indices have the dimensions of [P], where [P] is the basic dimensional unit in bibliometrics which is the unit publication or paper. The newly introduced Euclidean index, based on the Euclidean length of the citation vector has the dimensions $[P^{3/2}]$. An empirical example is used to illustrate the concepts.

**Introduction**

Evaluative and descriptive bibliometrics provide a quantitative focus on citations and/or publications (Andersen 2016). At the simplest level of aggregation, we study the performance of an individual scholar. At this micro-level, a controversial usage of indicators is to perform ranking and hard impact assessment to inform critical decisions about funding, promotion and tenure and the allocation of billions of dollars of research funding (Perry & Reny 2016). There is therefore a pressing need that indices (e.g., the *h*-index) go beyond heuristic rules of thumb and instead are founded on axiomatic principles. Five natural properties are considered and Perry & Reny (2016) propose a unique new index, the Euclidean index $i_E$, the Euclidean length of an individual's citation list. In this paper we shall show that following these five rules is not sufficient for robustness. There is a need to introduce one more requirement, that of dimensional homogeneity or consistency. We show that $i_E$ is not dimensionally commensurable with other citation indices like the impact $i$, or the $h$-index or many of its variants.

**Dimensional homogeneity**

Dimensional homogeneity is a well known principle in theoretical physics and engineering analysis. It requires that an equation must have quantities of same units on both sides. An equation is meaningful only it is homogeneous, with equality being applied between quantities of different nature. [M], [L] and [T] are the first such fundamental units to be encountered in physics, where they stand for the units of mass, length and time. Velocity or speed combines length and time as [L]/[T] = [$LT^{-1}$]. Momentum will be [M][L]/[T] = [$MLT^{-1}$], etc.

Extending this idea to bibliometrics, the basic dimensional unit in bibliometrics is the minimum unit of publication, namely the paper, say [P]. This has the same role as [M], [L] and [T] in physics. Leydesdorff (2009) and Bollen *et al.* (2009) have identified size and impact as the main categories in which the majority of bibliometric indicators can be arranged into (Andersen 2016). Size is measured as the number of papers $P$ (a numerical quantity) and its corresponding dimensional unit, in this case [P]. Impact is derived from the impact of all the $P$ papers in the portfolio. Thus if the k-th paper has $c_k$ citations, this means that this paper has been cited by $c_k$ papers. This is also the impact $i_k$ of the k-th paper. Here, $c_k$ or $i_k$ is the numerical quantity and the fundamental unit is again [P]. The total citations $C = \Sigma c_k$ then has the units [$P^2$] since the individual impact of each paper is summed over the total number of papers in the portfolio. The specific impact $i = C/P$ of the portfolio also has the units [P].

The best-known bibliometric indicator beyond the count of papers $P$, the impact $i$ or the total citation count $C$ is the $h$-index (Hirsch 2005). A scholar's $h$-index is the number, $h$, of his/her papers that each have at least $h$ citations. Fortuitously, this definition makes the $h$-index commensurable with $P$ and $i$; i.e. $h$ has the same dimensions as number of papers and the impact of the papers. Most of the variants of the $h$-index, such as the $g$-index (Egghe 2006a,b) have the same dimension and can be directly compared to each other. If indeed they had different dimensions, they are incommensurable and cannot be directly compared. In this paper we shall show that the newly proposed Euclidean index $i_E$ has a different dimension and so is not an alternative to nor can be compared to any of the other *h*-type indices.

**The three-dimensional nature of a citation distribution**

Leydesdorff (2009) and Bollen et al. (2009) see bibliometrics through a two-dimensional prism – quantity/size and impact (which can be interpreted as a proxy for quality or excellence) are the main categories in which most of bibliometric indicators can be arranged into (Andersen 2016). Prathap (2011a,b), proposed that comparative evaluation needs at least three dimensions: quantity/size, quality/excellence and consistency/balance or evenness. The quality-quantity-consistency parameter space leads to the evolution of second order indicators for any portfolio of papers (Prathap 2014a,b).

For any portfolio of publications, the total number of papers or articles, $P$, and the total number of citations, $C$, are often taken as indicators or proxy measures for the size of output of a group and the impact of its published research respectively (Katz 2005). The total impact, $C$, is size-dependent, and a specific impact, $i$, defined as $C/P$ is size-independent. The journal impact factor was defined in such a manner as a size-independent indicator to select journals for inclusion in the Science Citation Index. It was not originally intended to be a direct measure or proxy of quality (Pendlebury and Adams 2012), but since then has been accepted as a proxy or indirect measure of the quality or scholarly influence of a journal in a size-independent manner. In the same way, the scientific output of an individual or an entity can be measured using the following three-dimensional parameter space:

Quantity: No of papers/articles $P$ published during a prescribed publication window. This is a size-dependent proxy.

Quality: The impact $i$ computed as $C/P$ where $C$ is the number of citations during a prescribed citation window of all the articles $P$. Note that the definition of $i$ needs two distinct windows to be identified, the publication window and the citations window. The famous JIF is based on the use of a publication window of two years immediately preceding a single year citation window (Garfield 1955, 1999, 2005; Pendlebury and Adams 2012). This is a size-independent proxy.

Once the quantity $P$ and quality $i$ parameters are defined, it is possible to postulate the following sequence of indicators of performance (Prathap 2011b):

Zeroth order indicator: $P = i^0 P$
First order indicator: $C = i^1 P$
Second order indicator: $X = i^2 P = i^1 C$.

Thus both $P$ and $C$ serve as indicators of performance in their respective ways. Following Leydesdorff & Bornmann (2011), one can think of $C = iP$ as the first order integrated indicator for performance. Prathap (2011a,b) showed that the indicator $X = i^2 P$ can be thought of as a second order integrated indicator of performance. Since $X$ gives greater emphasis to quality than $C$ does, it is expected to be a better indicator or proxy of performance. Given the citation sequence $c_k$ of a portfolio of $P$ papers, this paradigm then leads to a trinity of second order terms (Prathap 2011a,b):

$X = i^2 P$
$E = \sum c_k^2$
$S = \sum (c_k - i)^2 = E - X$

where

$P = \sum 1$
$C = \sum c_k$
$i = C/P$.

In the foregoing, $X$ is exergy, $E$ is energy and $S$ is entropy. It is easy to see that $X$, $E$ and $S$ have the units $[P^3]$.

Hirsch (2005) requires citation sequences to be re-arranged in a monotonically decreasing order. Highly cited articles are usually found in a small core, implying a possible huge variation in the quality of each paper in the publication set. Two different portfolios can have the same *C*, and one could have achieved this with far fewer papers, with a higher quality of overall performance, or with the same number of papers (i.e., same quality) but a higher degree of consistency. Thus, *C* by itself, which is a first-order indicator may not be the last word on the measurement of performance. The product $X = iC = i^2 P$ is a robust second-order performance indicator, and is arguably a better proxy for performance (Prathap 2011a,b). Apart from *X*, an additional indicator *E* also appears as a second-order indicator as above. The coexistence of *X* (exergy) and *E* (energy) allows us to introduce a third attribute that is neither quantity nor quality. In a bibliometric context, the appellation "consistency" may be more meaningful. The simple ratio of *X* to *E* can be viewed as the third component of performance, namely, the consistency term $\eta = X/E = (C^2/P)/\Sigma c_k^2$. Perfect consistency ($\eta = 1$, i.e., when $X=E$) is a case of absolutely uniform performance (i.e. entropy *S* is zero); that is, all papers in the set have the same number of citations, $c_k = c$. The greater the skew, the larger is the concentration of the best work in a very few papers of extraordinary impact. The inverse of consistency thus becomes a measure of concentration. *X* by itself is a proxy of performance that ignores the actual distribution of the citations over the publication set. The ratio *X/E*, denoted as $\eta$, which is now dimensionless, takes into account the variability in the citation distribution of a portfolio of papers. It is important to emphasize again that this ratio is identical to evenness or balance in ecology, and also serves as an inverse measure of concentration, a term used by economists (Zhang et al. 2016).

**The Euclidean index**

Perry & Reny (2016) used five axioms or "basic properties", which they considered crucial for an indicator of an individual's citation impact. These are monotonicity, independence, depth relevance, scale invariance and directional consistency. They proposed a unique new index, the Euclidean index $i_E$, the Euclidean length of an individual's citation list. In terms of the foregoing nomenclature, it is given simply by $i_E = \sqrt{E}$. Immediately we see that the units for the Euclidean length is [$P^{3/2}$]. It is therefore not commensurable with *i* or *h* or any of the *h*-variants all of which have the units of [P]. We also notice that $i_E$ cannot capture the large skew in any citation distribution (that is the third dimension of consistency or evenness).

**An empirical test for dimensionality – leading authors in polymer solar cells research**

Prathap (2014c) gives typical citation indices for leading authors in an emerging area of research, namely polymer solar cells from the three-dimensional point of view. In Table 1, we re-interpret the original Table 3 of Prathap (2014c) now introducing the Euclidean index into the analysis. We see little correlation between the primary dimensions $P$, $i$ and $\eta$, showing that they are indeed orthogonal and therefore independent dimensions. Also shown in the table is the *z*-index defined as $(\eta i^2 P)^{1/3}$ which was introduced by Prathap (2014b) as an *h*-type index that is three-dimensional in nature. It has the dimensions of $[P]$.

To visualize the exponential relationships between these indices the results from Table 1 are plotted on logarithmic scales in Figure 1. We see very clearly that while the *z*-index and *h*-index scale identically as $[P]$, the Euclidean index scales as $[P^{3/2}]$ and C scales as $[P^2]$.

**Concluding remarks**

We show the importance of dimensional analysis and dimensional homogeneity in bibliometric analysis. It is seen that while most *h*-type indices have the dimensions of $[P]$, the newly introduced Euclidean index has the dimensions $[P^{3/2}]$. It is not enough to have an axiomatic basis for designing an indicator; it is necessary to examine for dimensional homogeneity to ensure that it is commensurable with other similar indicators.

Perry, M., & Reny, P. J. (2016). How To Count Citations If You Must. *American Economic Review*, 106 , 2722-2741. doi:10.1257/aer.20140850.

Prathap, G. (2011a). The Energy–Exergy–Entropy (or EEE) sequences in bibliometric assessment. *Scientometrics*, 87, 515–524.

Prathap, G. (2011b). Quasity, when quantity has a quality all of its own—toward a theory of performance. *Scientometrics, 88, 555–562*.

Prathap, G. (2014a). Quantity, Quality, and Consistency as Bibliometric Indicators. *Journal of the American Society for Information Science and Technology*, 65(1), 214.

Prathap, G. (2014b). The Zynergy-Index and the Formula for the h-Index. *Journal of the American Society for Information Science and Technology*, 65(2), 426-427.

Prathap, G. (2014c). A three-dimensional bibliometric evaluation of research in polymer solar cells. *Scientometrics*, 101, 889-898. doi:10.1007/s11192-014-1346-z

Zhang, L., Rousseau, R., & Glänzel, W. (2016). Diversity of References as an Indicator of the Interdisciplinarity of Journals: Taking Similarity Between Subject Fields Into Account *Journal of the Association for Information Science and Technology*, 67(5):1257–1265.8

Table 1 The leading authors in polymer solar cells research ranked according to the default quantity parameter $P$

| AUTHORS | P | i | η | h | z | $i_E$ | C |
|---|---|---|---|---|---|---|---|
| Dimensions | [P] | [P] | nil | [P] | [P] | [$P^{3/2}$] | [$P^2$] |
| LI YF | 142 | 33.25 | 0.20 | 34 | 31.41 | 891.42 | 4721 |
| KREBS FC | 96 | 73.05 | 0.24 | 41 | 49.69 | 1462.71 | 7013 |
| YANG Y | 78 | 128.65 | 0.12 | 37 | 53.69 | 3281.34 | 10035 |
| JANSSEN RAJ | 56 | 53.32 | 0.17 | 24 | 30.13 | 962.81 | 2986 |
| HOU JH | 45 | 99.89 | 0.17 | 21 | 42.15 | 1640.71 | 4495 |
| JEN AKY | 45 | 48.71 | 0.42 | 23 | 35.51 | 504.50 | 2192 |
| CAO Y | 44 | 38.73 | 0.18 | 15 | 22.97 | 599.26 | 1704 |
| KIM H | 44 | 9.55 | 0.26 | 11 | 10.18 | 123.38 | 420 |
| YIP HL | 44 | 49.82 | 0.43 | 23 | 36.05 | 504.50 | 2192 |
| ZHANG FL | 44 | 62.32 | 0.32 | 23 | 37.86 | 733.40 | 2742 |
| CORRELATION | P | i | η | h | z | $i_E$ | C |
| P | 1.00 | 0.04 | -0.35 | 0.74 | 0.27 | 0.29 | 0.53 |
| i | 0.04 | 1.00 | -0.41 | 0.55 | 0.88 | 0.92 | 0.83 |
| η | -0.35 | -0.41 | 1.00 | -0.24 | -0.14 | -0.60 | -0.52 |
| h | 0.74 | 0.55 | -0.24 | 1.00 | 0.81 | 0.65 | 0.86 |
| z | 0.27 | 0.88 | -0.14 | 0.81 | 1.00 | 0.78 | 0.85 |
| $i_E$ | 0.29 | 0.92 | -0.60 | 0.65 | 0.78 | 1.00 | 0.94 |
| C | 0.53 | 0.83 | -0.52 | 0.86 | 0.85 | 0.94 | 1.00 |

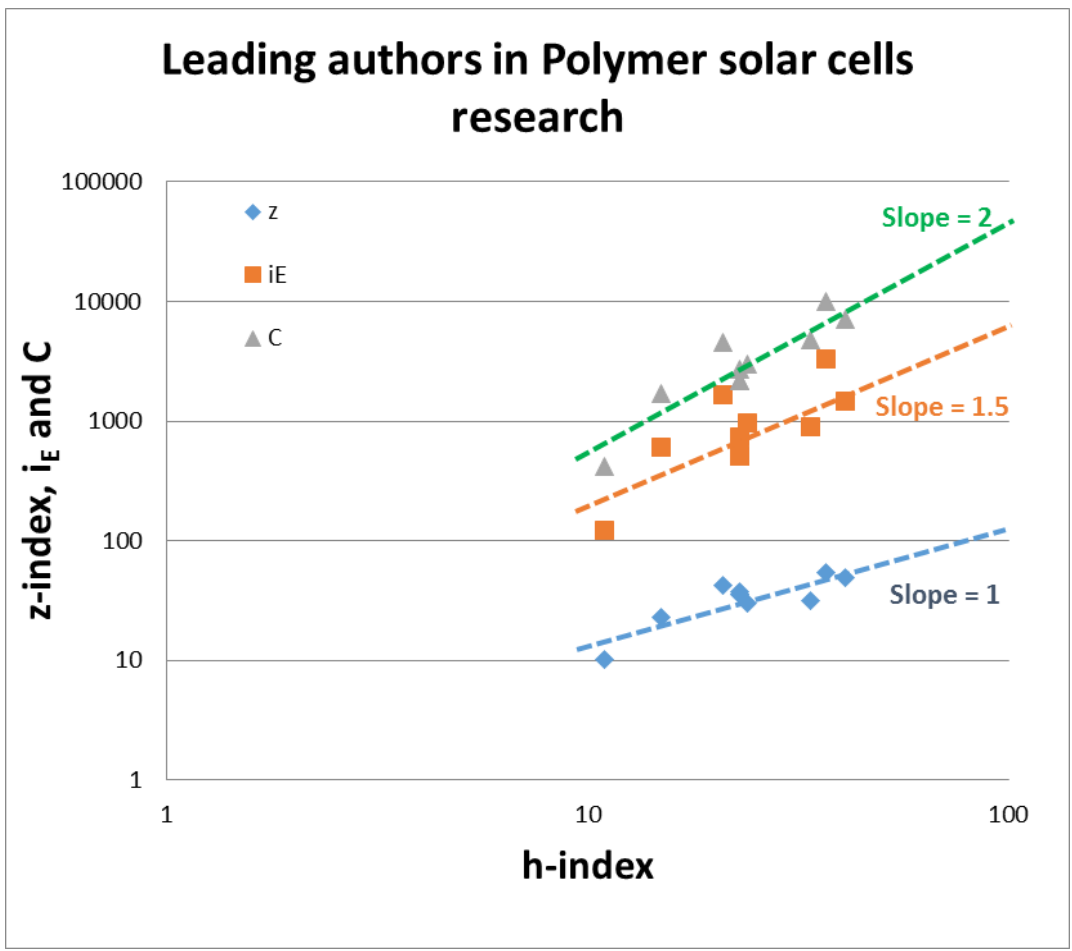

Figure 1. The dimensional relationship between various citation indices